\documentclass[11pt,letterpaper]{article}
\usepackage{comment}
\usepackage{jheppub}
\usepackage{microtype,color}
\usepackage[T1]{fontenc}
\usepackage{complexity}
\usepackage{tikz,pgfplots}

\DeclareMathOperator{\Tr}{Tr}

\title{Bootstrapping Lattice Vacua}
\author{Scott Lawrence}
\affiliation{Department of Physics, University of Colorado, Boulder, CO 80309, USA}
\emailAdd{scott.lawrence-1@colorado.edu}

\abstract{This paper demonstrates the application of semidefinite programming to lattice field theories, showcasing spin chains and lattice scalar field theory. Requiring expectation values of manifestly positive semi-definite operators to be non-negative results in a lower bound on the ground-state energy of any quantum mechanical system, which can be made arbitrarily tight for systems described by finite-dimensional Hilbert spaces. Such bounds can be obtained directly in the infinite-volume limit. The process of optimizing these lower bounds also yields estimates for a chosen set of expectation values in the ground state.}
\toccontinuoustrue

\begin{document}
\maketitle \flushbottom

\newpage

\section{Introduction}\label{sec:introduction}

Lattice Monte Carlo methods have been enormously successful in providing thermal and ground state information about field theories, most notably quantum chromodynamics. However, it is always useful to have more than one family of algorithms available. For example, lattice methods famously struggle with insufficiently symmetric fermionic systems, including the Hubbard model away from half-filling and relativistic systems at non-vanishing fermion density, due to the fermion sign problem, which provably lacks a general solution~\cite{troyer2005computational}. The purpose of this paper is to introduce a new method for estimating ground state energies and expectation values, by recasting the problem as a \emph{semi-definite program} (SDP).

A semi-definite program can be defined as follows. Given a Hermitian matrix $C$, we wish to minimize the function $\Tr C^\dag X$ over the space of positive semi-definite matrices $X \succeq 0$, subject to some linear constraints on the matrix elements of $X$. We can describe the $m$ linear constraints by $m$ matrices $A_i$ and vectors $b_i$, and demanding that $\Tr A_i^\dag X = b_i$. In short, then, an SDP consists of a Hermitian matrix $C$, and $m$ matrix-vector pairs $(A_i, b_i)$, and is solved by minimizing
\begin{equation}\label{eq:sdp}
\langle C,X\rangle\text{ subject to }\langle A_i,X\rangle = b_i
\text,
\end{equation}where we have introduced the inner product notation $\langle M_1,M_2\rangle \equiv \Tr M_1^\dag M_2$.

In recent years, many physical calculations have been profitably recast as SDPs. Perhaps the most famous is the conformal bootstrap~\cite{Rattazzi:2008pe}, which has yielded remarkably precise estimates for the critical exponents of the Ising model in three dimensions~\cite{Kos:2016ysd} (with impressive results for many other CFTs besides) --- see~\cite{Poland:2018epd} for a review of these methods. Quantum field theories without conformal invariance can be treated with a similar philosophy, resulting in the S-matrix bootstrap~\cite{chew1961s}, although only recently have SDPs been used to solve the resulting equations~\cite{He:2018uxa,Caron-Huot:2020cmc}. Similar positivity-based methods have been proposed to study gauge theories~\cite{Anderson:2016rcw}.

Most recently, SDP-based methods have been successfully applied to a variety of quantum mechanical systems, typically yielding eigenenergies to high precision~\cite{Han:2020bkb,Berenstein:2021dyf,Berenstein:2021loy}. In this paper, we will use the Hamiltonian formalism of lattice field theory to adapt those approaches to obtain lower bounds on the ground state energy in field theories, as well as estimates for expectation values in the ground state. In combination with variational methods, this results in an estimate of the ground state energy with precisely known systematics. An early incarnation of many of the ideas in this paper is found in~\cite{barthel2012solving}, including an application to spin systems and the Hubbard model.

The remainder of the paper is organized as follows. We begin in section~\ref{sec:anharmonic} by introducing a simplified variant of the quantum mechanical bootstrap method studied by~\cite{Han:2020bkb,Berenstein:2021dyf,Berenstein:2021loy}, designed to bound only the ground state energy (from below). This method is demonstrated using the anharmonic oscillator as a case study. Section~\ref{sec:spins} applies the method to spin chains. In the same section, we establish that for finite systems, the bounds given by the bootstrap eventually converge to the true ground-state energy. In section~\ref{sec:greedy}, a faster algorithm is constructed by disregarding constraints that are not improving the bound. A modified algorithm that works directly in the infinite-volume limit is described in section~\ref{sec:limit}, and we show how this algorithm, conjoined with standard tensor-network-based variational methods, yields a tight estimate of the ground state energy. With that groundwork laid, section~\ref{sec:scalar} tackles a field theory of one scalar field in one spatial dimension. Finally, we conclude in section~\ref{sec:discussion} by discussing likely future steps.

All code used for this paper is available online~\cite{code}. SDPs are solved using the MOSEK toolkit~\cite{mosek}.

\section{Anharmonic oscillator}\label{sec:anharmonic}

In this section we detail the general approach we take to constraining the ground state energy, applied to a system of a single particle in a potential well:
\begin{equation}\label{eq:H_osc}
H_{\mathrm{osc}} = \frac 1 2 p^2 + \frac {m^2} 2 x^2 + \frac\lambda 4 x^4
\text.
\end{equation}

Related methods were applied to this system in~\cite{Han:2020bkb,Berenstein:2021dyf} in order to obtain bounds on the permitted eigenenergies. Here we will describe a simpler (and more numerically efficient) method, at the cost of only obtaining information about the ground state. As the ground state of a field theory is generally more interesting than any other eigenstate, this will be a worthwhile trade-off in future sections.

The key constraint is that for any operator $\mathcal O$, a corresponding expectation value must be real and non-negative: $\langle \mathcal O^\dag \mathcal O\rangle \ge 0$. Choosing a basis of operator $\mathcal O_i$, this constraint implies that the matrix
\begin{equation}
X_{ij} \equiv \langle \mathcal O_i^\dag \mathcal O_j\rangle
\end{equation}
must be positive semi-definite, as for any vector $v$ we have $v^\dag X v = \langle (v \cdot \mathcal O)^\dag (v \cdot \mathcal O) \rangle \ge 0$. In this section we will work with the basis $\{1, x, x^2, \ldots, p, p^2, \ldots\}$.

The matrix $C$, which defines the objective function, is obtained from the Hamiltonian. In the case of (\ref{eq:H_osc}), we have
\begin{equation}
C_{(p)(p)} = \frac 1 2\text{, \;} C_{(x)(x)} = \frac {m^2} 2\text{, \;and \;}C_{(x^2)(x^2)} = \frac\lambda 4
\text,
\end{equation}
with all other matrix elements equal to $0$. This picks out the expectation value $\langle C,X\rangle = \langle H \rangle$ given a set of expectation values represented in $X$.

We will assume that the basis of operators $\mathcal O_i$ is chosen to be linearly independent. Even so, not all matrix elements of $X$ are independent. As an example, $\langle x p \rangle$ and $\langle p x\rangle$ are of course related as a result of the canonical commutation relation $[x,p] = i$. Each matrix element of $X$ can be re-written as a linear combination of operators in a standardized order with all $x$-operators preceding all $p$-operators\footnote{Had we chosen creation and annihilation operators as the basis, it would be most convenient to work with normal-ordered operators.}. This procedure defines the linear constraints that must be imposed on the matrix $X$.

This completes the description of the SDP to be solved. Nothing in this SDP singles out the ground state. In fact, any set of expectation values achievable by any pure or mixed state will obey the constraints described above. Nevertheless, minimizing the expectation value of the Hamiltonian yields the ground state energy, and expectation values in the ground state can be read off from the optimal $X$.

In principle, we could impose more constraints. In~\cite{Berenstein:2021dyf}, the fact that $\langle H \mathcal O\rangle = E\langle \mathcal O\rangle$ for all $\mathcal O$ was used (representing a restriction to eigenstates), but in our case this is inconvenient, as it is nonlinear. A slightly weaker option, which only restricts the state to the space of density matrices that commute with the Hamiltonian, is to require $\langle [H,\mathcal O]\rangle = 0$. However, in practice including this constraint does not much improve the estimates, and it is not used in what follows.

Of course, the SDP as described above cannot be solved on any computer --- the matrices involved are infinite-dimensional, so a truncation is first needed. To truncate, we chose any incomplete basis, such as $\{1,x,x^2,p\}$, construct the corresponding SDP, and solve. Note that this procedure only removes constraints, and therefore the optimal value of $\langle C,X\rangle$ is guaranteed to be lower than the true ground state energy. In this way, the truncated SDP acts as a natural companion to variational methods, as one yields a lower bound and the other an upper.

\begin{figure}
\hfil
\includegraphics[width=0.45\textwidth]{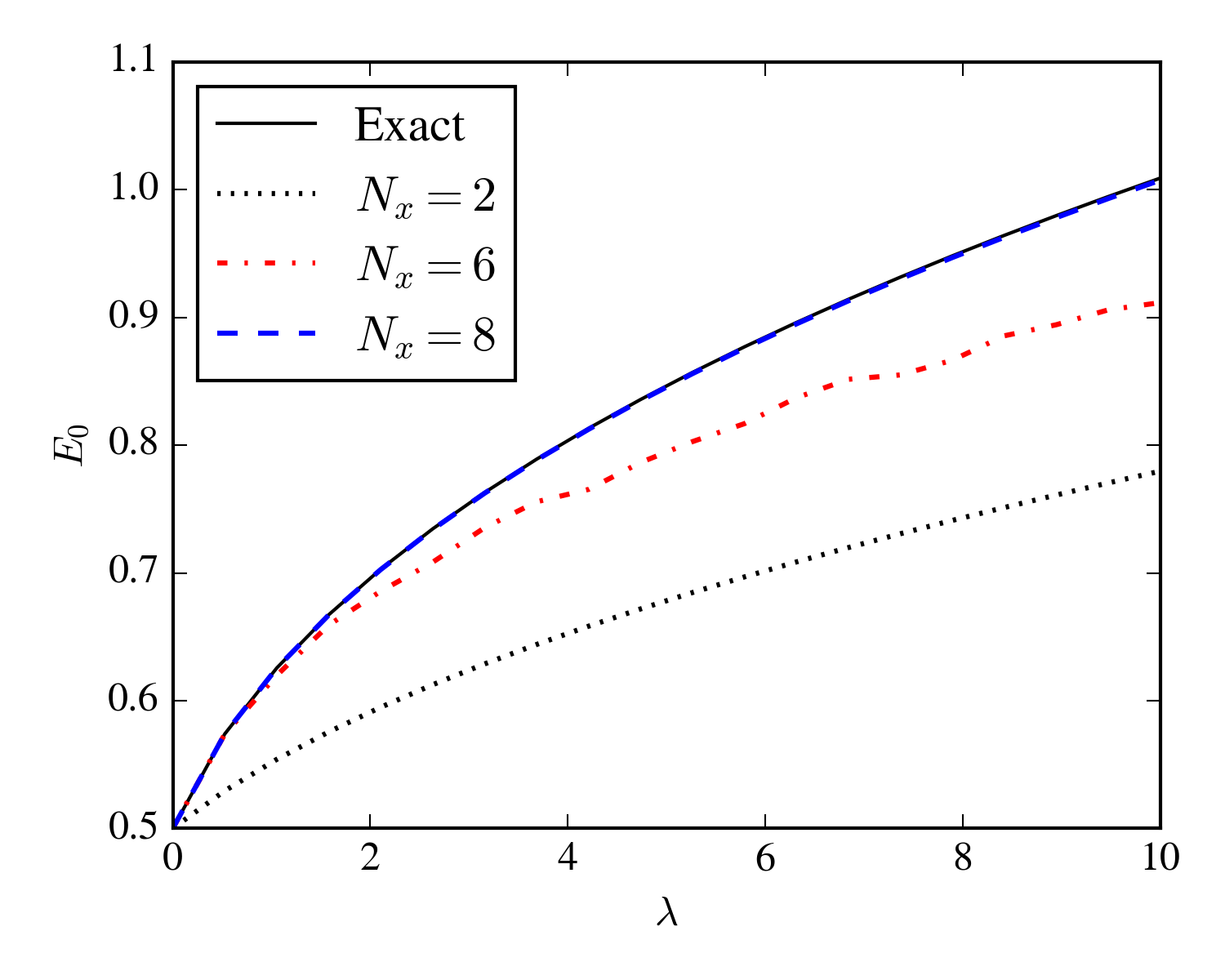}
\hfil
\includegraphics[width=0.45\textwidth]{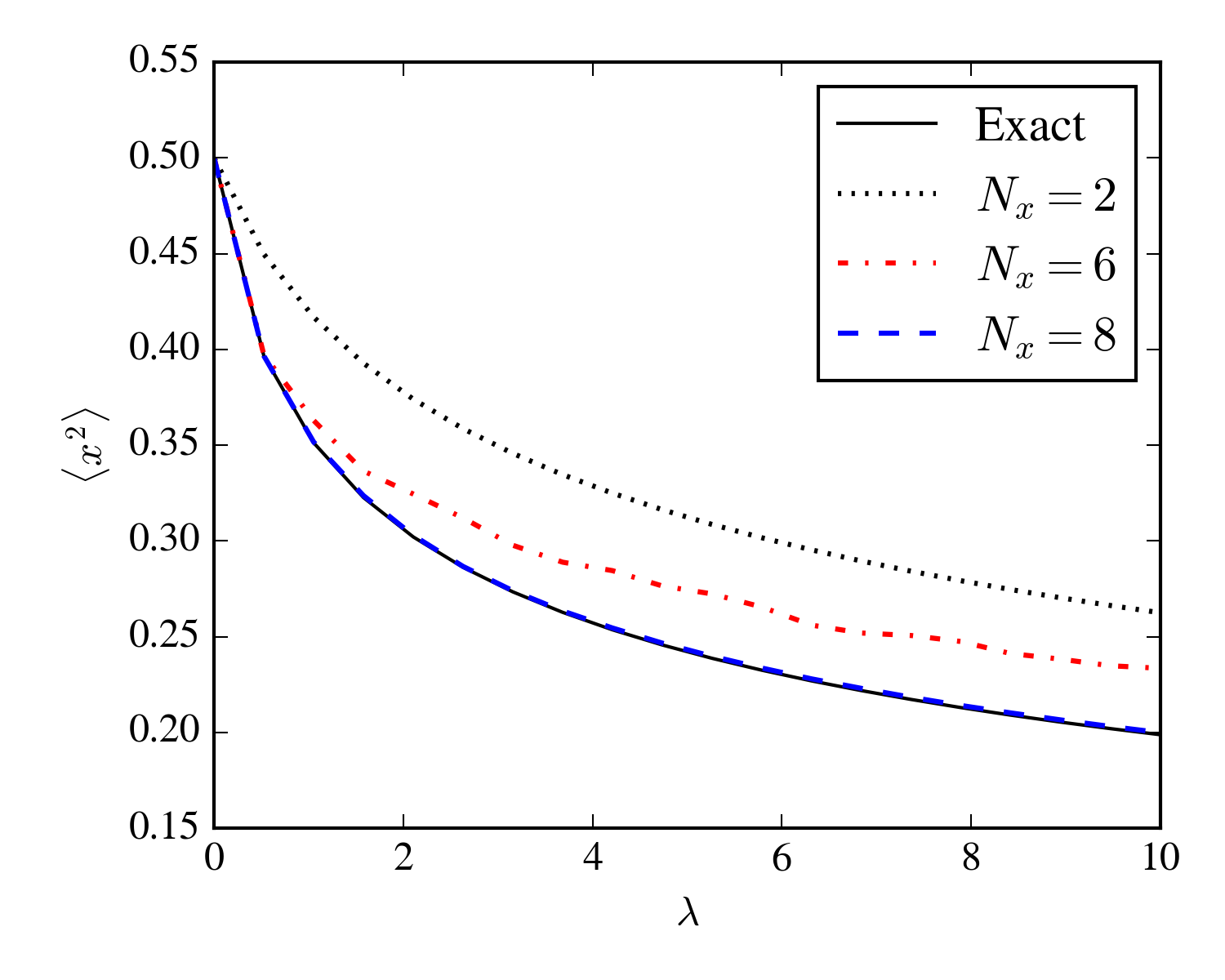}
\hfil
\caption{\label{fig:aho}Demonstration of the SDP method on the anharmonic oscillator (\ref{eq:H_osc}). On the left, the ground state energy is estimated with three different truncations, the last of which is within $1\%$ of the exact value. On the right, the expectation value of $x^2$ in the ground state is approximated.}
\end{figure}

To demonstrate, we estimate the ground state energy $E_0$ and the ground state expectation value $\langle x^2\rangle$ for a range of couplings $\lambda \in [0,10]$. The basis of operators used is $\{1,p,x,x^2,\ldots,x^{N_x}\}$. We find that including operators $p^2$ and higher does not improve the estimate.

The left panel of figure~\ref{fig:aho} shows the bounds obtained for $m=1$ and varying $\lambda$, as the size of the truncated basis is increased. The true ground-state energy (obtained by diagonalizing the Hamiltonian in truncated harmonic oscillator basis) is shown for comparison. The convergence to the true value is, as reported in~\cite{Berenstein:2021dyf}, at least exponentially quick.

The right-hand panel of figure~\ref{fig:aho} shows the estimate of the expectation value $\langle x^2\rangle$, again as a function of coupling $\lambda$ and for the same truncations that were used in the left panel. As with variational methods, these expectation values no longer represent rigorous bounds, but still converge to the true expectation value as the truncation is lifted.

In practice, lifting $N_x$ to be too high causes numerical instabilities in the SDP solver, meaning that high-precision estimates of the ground-state energy cannot be obtained with this method. These instabilities kick in near $N_x \sim 14$, and limit the calculation to estimates no better than one part in $10^3$. Despite the appearance of figure~\ref{fig:aho}, the result with a truncation of $N_x = 8$ is not exact\footnote{This can be verified by computing --- with the aid of a multi-precision math library --- a set of expectation values for the anharmonic oscillator via exact diagonalization. The matrix of expectation values $X$, constructed from the operators in the $N_x = 8$ truncation, has a smallest eigenvalue of order $10^{-9}$. This indicates that the true ground-state energy is not optimal for the $N_x=8$ SDP, and therefore that higher truncations are needed to obtain an exact result, even at arbitrary precision.}.

\subsection{Sum-of-squares}
The solution of the harmonic oscillator begins by noting that the Hamiltonian can be rewritten as an operator times its Hermitian conjugate: $x^2 + p^2 = \frac 1 2 + a^\dag a$. This immediately yields a lower bound (which happens to be tight) on the ground state energy, as $a^\dag a$ is manifestly positive semi-definite. As we will now see, the solution of the SDP above can be rephrased as a generalization of this procedure.

The SDP (\ref{eq:sdp}) has a dual formulation: instead of minimizing $\langle C,X\rangle$ over a space of positive semi-definite matrices, we will maximize $b \cdot y$ over all $y \in \mathbb R^m$, subject only to the constraint that $C - y \cdot A$ be positive semi-definite. For any vector $y$ obeying $C \succeq y\cdot A$, $b \cdot y$ is a lower bound on $\langle C,X\rangle$:
\begin{equation}
0 \le \langle C - y \cdot A,X\rangle
= \langle C,X\rangle - y \cdot \langle A,X\rangle
= \langle C,X\rangle - y \cdot b
\text.
\end{equation}
Maximizing $b \cdot y$ thus yields the tightest possible lower bound on the ground state energy.

The vector $y$ has a nice physical interpretation as a recipe for rewriting the Hamiltonian as a sum of squared operators. Recall first that $C$ specifies the Hamiltonian as a linear combination of expectation values (i.e.\ matrix elements of $X$). Moreover, each $A_i$ specifies a linear combination of expectation values that is required to be equal to the constant $b_i$; thus $y\cdot A$ describes an expectation value that, although apparently nontrivial when written out explicitly, in fact equals $y \cdot b$. Finally, if a matrix $M$ is positive semi-definite, then it can be taken to represent a positive semi-definite operator $H_M$ on the Hilbert space. To see this, note that for every state $|\psi\rangle$, there is a corresponding positive semi-definite $X_\psi$. The expectation value of the operator corresponding to $M$ is given by $\langle\psi|H_M|\psi\rangle = \Tr M^\dag X_\psi$; since both $M$ and $X_\psi$ are positive semi-definite, we have $\langle\psi|H_M|\psi\rangle$. This holds for every $\psi$, and so we see that $H_M$ is indeed positive semi-definite.

Putting this all together, we see that finding a positive semi-definite matrix $C - y\cdot A$ is equivalent to writing the Hamiltonian as a constant plus a positive semi-definite operator. Diagonalizing the matrix $C - y\cdot A$ puts this operator in a form that is manifestly positive semi-definite, as a sum of Hermitian-squared operators.

Let us see how this method applies directly to the anharmonic oscillator Hamiltonian (\ref{eq:H_osc}). With the minimal $N_x = 4$ truncation, the matrix $X$ of expectation values is given by
\begin{equation}
X = \left(\begin{matrix}
1 & \langle p \rangle & \langle x \rangle & \langle x^2 \rangle\\
\langle p \rangle & \langle p^2 \rangle & \langle xp \rangle - i & \langle x^2 p\rangle - 2 i \langle x \rangle \\
\langle x \rangle & \langle xp \rangle & \langle x^2 \rangle & \langle x^3 \rangle \\
\langle x^2 \rangle & \langle x^2 p \rangle & \langle x^3 \rangle & \langle x^4 \rangle
\end{matrix}\right)\text{, \;and \;}
C = \left(
\begin{matrix}
0 & 0 & 0 & 0\\
0 & \frac 1 2 & 0 & 0\\
0 & 0 & \frac{m^2}{2} & 0\\
0 & 0 & 0 & \frac{\lambda}4\\
\end{matrix}
\right)
\end{equation}
represents the Hamiltonian according to $\langle H \rangle = \Tr C X$. Now, $C$ is already diagonal, because (\ref{eq:H_osc}) already expresses the Hamiltonian as a sum of Hermitian-squared operators. This form of the Hamiltonian corresponds to the trivial bound $\langle H \rangle \ge 0$. A stricter bound can be obtained by adding to $C$ terms corresponding to a combination of expectation values that must be equal to a negative constant. For example, let us take
\begin{equation}
C' = \left(
\begin{matrix}
0 & 0 & 0 & 0\\
0 & \frac 1 2 & ia & 0\\
0 & -ia & \frac{m^2}{2} & 0\\
0 & 0 & 0 & \frac{\lambda}4\\
\end{matrix}
\right)
\end{equation}
for real $a \ge 0$. This corresponds to adding a term $ia(xp - px) = -a$ to the Hamiltonian. Diagonalizing the inner two-by-two block reveals that the largest $a$ can be, while keeping $C'$ positive semi-definite, is $\frac m 2$, which is of course the true ground-state energy when $\lambda = 0$, and a lower bound for all non-negative $\lambda$. Furthermore, at $a = \frac m 2$, the eigenvector with non-negative eigenvalue corresponds to the operator $(m x + i p)$, yielding a new decomposition of the Hamiltonian:
\begin{equation}
H_{\mathrm{osc}} = \frac m 2+ \frac 1 2 (m x + i p) (m x - i p) + \frac\lambda 4 x^4
\text.
\end{equation}

Generalizations of this method go by the name sum-of-squares or non-commutative sum-of-squares; see~\cite{Hastings:2021ygw} for an application to the SYK model.


\section{Spin chain}\label{sec:spins}

Now consider a Heisenberg spin chain on $L$ sites. The Hamiltonian of this system is
\begin{equation}\label{eq:H_spins}
H_{\mathrm{spins}} = -\mu \sum_r \sigma_x(r) - \sum_i \left[J_i \sum_{\langle r r'\rangle} \sigma_i(r) \sigma_i(r')\right]\text,
\end{equation}
where the first sum is taken over all sites $r$, and the second over $i=x,y,z$ and all adjacent pairs of sites $(r,r')$. We will use periodic boundary conditions, so that sites $1$ and $L$ are considered to be adjacent. For suitably tuned parameters $\mu$ and $J_i$, this system exhibits a divergent correlation length (and sometimes Lorentz invariance), and is therefore described as a lattice field theory. For special values of the interaction strength $J_i$ and the external magnetic field $\mu$, this is an integrable system solvable by the Bethe ansatz. For example; with $J_x = J_y = 0$, the spin chain is equivalent (up to an extra term on the boundary) to a theory of free lattice fermions. To avoid such regimes, this section considers only the XYZ Heisenberg chain, with $J_z = 2 J_y = 3 J_x = 1$ and $\mu > 0$.

The first step in converting the Hamiltonian to an SDP is choosing a basis. A natural basis to use here is the set of length-$L$ Pauli strings; that is, assignments of one Pauli matrix (including the identity) to each lattice site. This basis, as with any complete basis, has $4^L$ elements, and is therefore impractical for any but very short spin chains\footnote{For contrast, direct diagonalization requires working only with a matrix of dimension $2^L$.}. A simple truncation is given by considering the $n$-point operators, for which the Pauli string contains only $n$ operators that are not the identity. Further reductions are obtained by limiting the separation of the operators, or by disregarding certain channels --- always at a potential loss to the tightness of the bound.

In place of the canonical commutation relation $[x,p] = i$, we now have the angular momentum relations
\begin{equation}
\sigma_x(r)\sigma_y(r) = i \sigma_z(r)\text{, }
\sigma_y(r)\sigma_z(r) = i \sigma_x(r)\text{, and }
\sigma_z(r)\sigma_x(r) = i \sigma_y(r)\text{. }
\end{equation}
Pauli operators at different sites of course commute: $[\sigma_i(r),\sigma_j(r')] = 0$. Using these relations, every operator can be expressed uniquely as a linear combination of Pauli strings. This procedure defines the linear relations between the matrix elements of $X$.

The rest of the formulation and solution of the SDP proceeds exactly as in the case of the anharmonic oscillator. The one-site case can be examined by hand, as there are only three operators in the basis. The matrix $X$ is written in terms of the three expectation values as
\begin{equation}
X = \left(\begin{matrix}
1 & \langle \sigma_x \rangle & \langle \sigma_y \rangle & \langle \sigma_z \rangle\\
\langle \sigma_x \rangle & 1 & i \langle \sigma_z\rangle & -i \langle \sigma_y \rangle\\
\langle \sigma_y \rangle & -i \langle \sigma_z\rangle & 1 & i \langle \sigma_x \rangle\\
\langle \sigma_z \rangle & i \langle\sigma_y\rangle & -i \langle \sigma_x \rangle & 1\\
\end{matrix}\right) \text.
\end{equation}
The task of the SDP is to minimize $\langle H \rangle = - \mu \langle \sigma_x\rangle$ while keeping $X$ positive semi-definite. The upper-left $2\times 2$ block of $X$ reveals the constraint $-1 \le \langle \sigma_x\rangle \le 1$. Taking the other two expectation values $\langle \sigma_y \rangle = \langle \sigma_z \rangle = 0$ to vanish, we see that $X$ can be made positive semi-definite for any $\langle \sigma_x\rangle$ in this range. Therefore, $\langle H \rangle$ is minimized by choosing $\langle \sigma_x\rangle = 1$, yielding the true ground state energy $E_0 = -\mu$. The optimal sum-of-squares formulation yields $H = \frac 1 2 (1 + \sigma_x)^2 - 1$, and the same bound.

\begin{figure}
\centering
\includegraphics[width=0.65\textwidth]{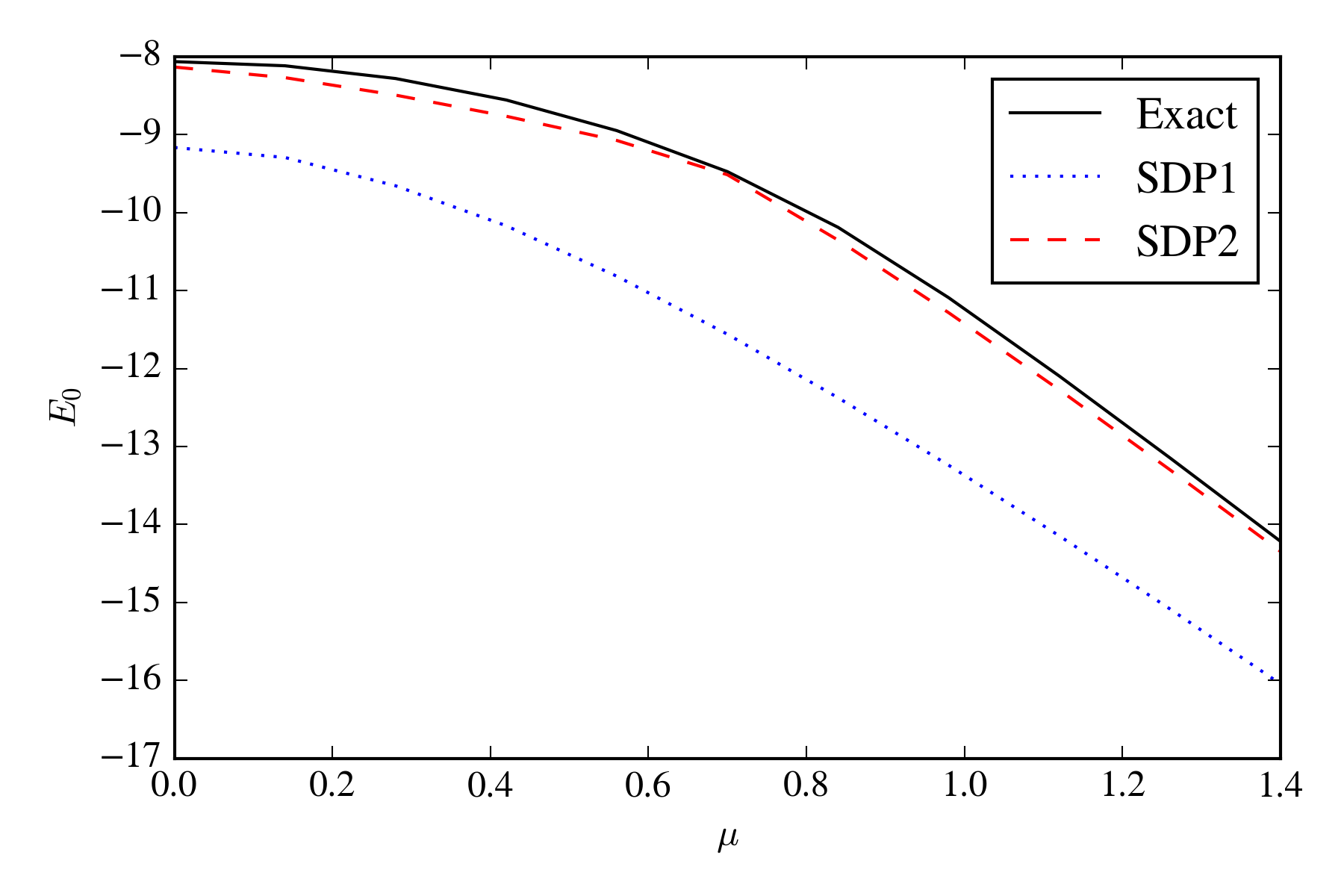}
\caption{\label{fig:spin}Ground state energy, exact and estimated, for the $8$-site periodic XYZ spin chain with Hamiltonian (\ref{eq:H_spins}). The operator bases used to compute the lower bounds are described in the text.}
\end{figure}

Returning to the full spin chain (\ref{eq:H_spins}), figure~\ref{fig:spin} shows the exact and estimated ground state energy for the XYZ spin chain described above, with $L=8$ sites. Two different operator bases are used. The first (``SDP1'' in the figure) consists of $\sigma_x(r)$ for every site $r$, as well as $\sigma_x(r)\sigma_x(r')$ and $\sigma_z(r)\sigma_z(r')$ for all adjacent pairs $(r,r')$, for a total of $3L = 24$ operators. The second, (``SDP2'') adds to this the $L$ operators $\sigma_y(r)\sigma_y(r')$; this basis has $4L = 32$ operators. Note that the first basis does not include every operator in the Hamiltonian; however, the matrix $X$ does contain (off of the diagonal) the missing $\sigma_y(r)\sigma_y(r')$ operators, which is sufficient to obtain a nontrivial bound.

Each expectation value included in the formulation of the SDP represents a linear functional on the space of density matrices $\rho$. A density matrix uniquely yields a complete set of expectation values. In the reverse direction, a complete set of expectation values uniquely constructs a Hermitian matrix $\rho$. Density matrices have the additional physical requirement that $\rho \succeq 0$, which is satisfied precisely when the matrix $X$ is itself positive semi-definite.
As a result, for any finite system, when all operators are included in the formulation of the SDP, a precise minimization is guaranteed to yield the true ground state energy.

In general, there is no reason to expect that any set of operators, short of the complete list of size $4^L$, yields the exact ground-state energy. In practice, of course, it may be hoped that the convergence is rapid, as is the case in figure~\ref{fig:spin}. However, in special cases, relatively small sets of operators yield an exact ground state energy as a bound. This was already seen in the previous section, where in the case of $\lambda = 0$, the operator basis $\{p, x\}$ immediately yields the exact ground-state energy $\frac m 2$. In general, a non-interacting theory will be subject to a similarly dramatic simplification. A non-interacting Hamiltonian is one which can be decomposed into a sum of terms $H = H_1 + H_2 + \cdots$, where $H_1$ and $H_2$ act on disjoint components of the Hilbert space. Typically, the number of terms in this decomposition is proportional to the volume. The ground state energy is given by the sum of ground state energies for each $H_i$, so a tight bound can be obtained by bounding each $H_i$ individually. This means that a sufficient basis for a tight bound scales linearly with the volume (excluding operators that couple different sectors of the theory), in place of the usual exponential growth.

At the beginning of this section was noted the fact that the Heisenberg spin chain with $J_x = J_y$, and open boundary conditions, is equivalent to a theory of free fermions. As a result, a small basis of operators --- for instance, the set of all one- and two-point functions --- is sufficient to obtain a tight bound. It is reasonable to conjecture that similar simplifications hold not just for free theories, but for all integrable theories.

\section{Greedy algorithm}\label{sec:greedy}

The optimization performed in solving an SDP requires repeated operations on a large matrix. If the truncated basis has $N$ elements, the minimization requires time $O(N^\alpha)$, typically with $\alpha \sim 3$. Although this is not prohibitive for reasonable $N$, it makes it desirable to be `economical' with basis elements, excluding those that are not contributing substantially to the bound.

The previous section used the crude heuristic that few-point functions would be more valuable in constraining the ground-state energy (which is itself a combination of one- and two-point functions) than many-point functions. This section describes an automated procedure that can be used to select valuable operators in the absence of further physical insight.

As alluded to in the previous section, a certain number of operators must be included in the basis no matter what: in order to obtain any bound at all, the matrix $X$ must depend on every operator that appears in the Hamiltonian. Therefore we will always begin with a minimal set of operators of size $3L$. These are $\sigma_x(r)$, $\sigma_y(r)\sigma_y(r')$, and $\sigma_z(r)\sigma_z(r')$. (We expect that $\sigma_y(r)\sigma_y(r')$ is more useful than $\sigma_x(r)\sigma_x(r')$ simply because $J_y > J_x$, and indeed this is readily confirmed by a quick numerical test.)

The optimized basis is constructed by adding one operator at a time to this starting basis. The operator is selected from a list of $K$ candidate operators, manually chosen in advance. The operator added is the one that raises the bound by the largest amount. This is decided by a naive search: each of the $K$ candidates is tentatively added to the basis, and the SDP solver is re-run to determine what bound is obtained with that operator included. This procedure is not guaranteed to yield the best possible bound for a basis of any given size; it merely serves as an automated heuristic for constructing a workably small basis from a much larger list of candidate elements. Because this algorithm is merely linear in $K$ (while approximately cubic in the chosen basis size $N$), a much larger set of candidates can be considered than could ever be included in the basis used to perform the computation.

\begin{figure}
\centering
\includegraphics[width=0.65\textwidth]{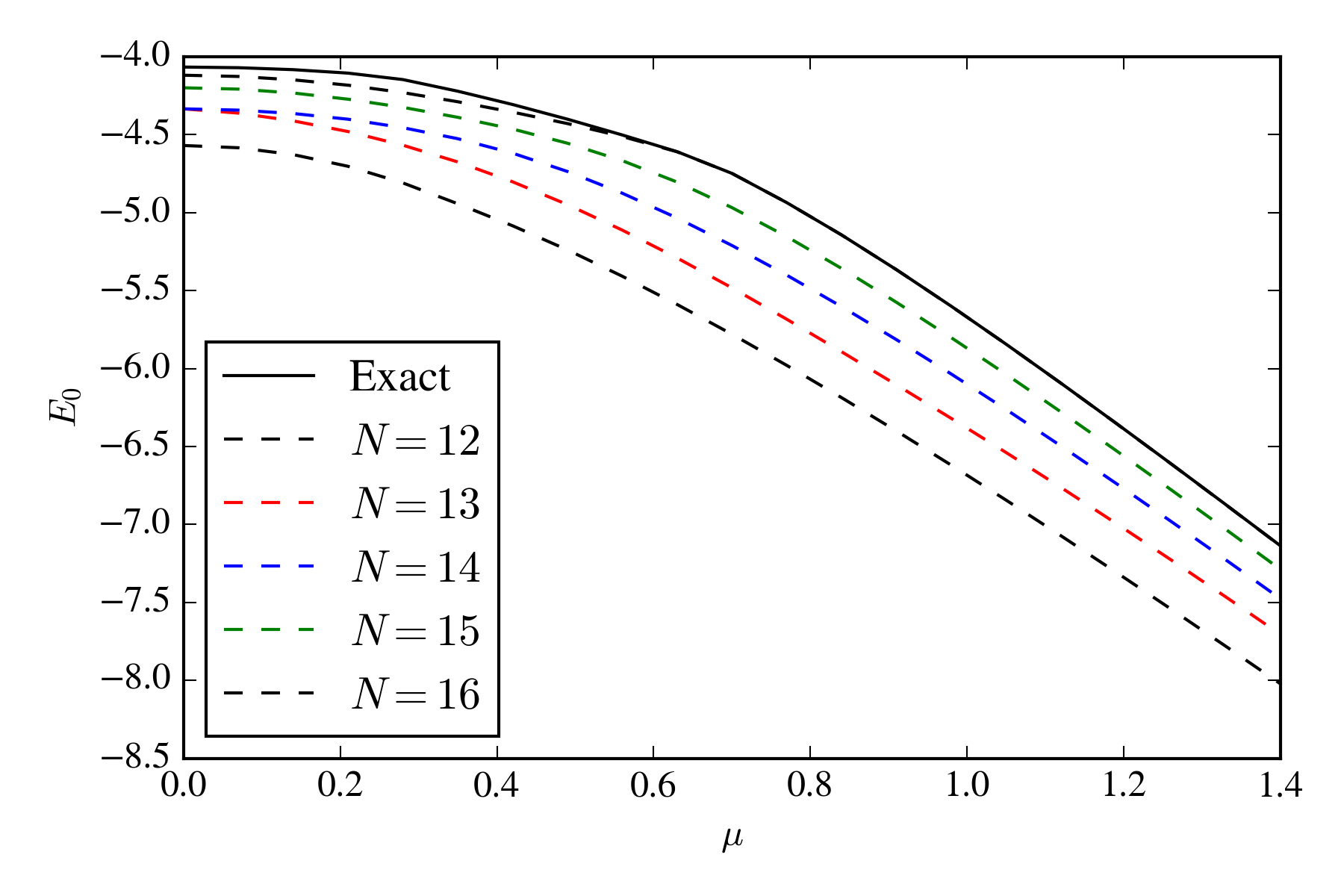}
\caption{\label{fig:greedy}Ground state energy, exact and estimated, for the $4$-site periodic XYZ spin chain with Hamiltonian (\ref{eq:H_spins}). Successive truncations, each with $N$ basis elements, are constructed by the algorithm described in the text. The truncation $N=12$ is equivalent to ``SDP1'' in figure~\ref{fig:spin}; ``SDP2'' has 16 basis elements, but yields a slightly inferior bound to the $N=16$ truncation constructed here.}
\end{figure}

Figure~\ref{fig:greedy} demonstrates this method on a $4$-site spin chain. The initial basis is taken to be ``SDP1'' from the previous section; that is, $\sigma_x(r)$ at every site, and $\sigma_x(r)\sigma_x(r')$ and $\sigma_z(r)\sigma_z(r')$ at all adjacent pairs of sites. The candidate list consists of all $1$- and $2$-point functions. The construction of an optimal basis is performed only once, at $\mu = 0.7$, and that basis is then used to compute a bound across the entire range of $\mu$ plotted. The initial basis has $12$ operators, and after $4$ additional operators have been added the bound is visually indistinguishable from the exact value for most values of $\mu$.

How can we characterize the performance of this method? In principle, we could test all possible bases of $N$ operators to determine what the best possible bound is, and see how close the greedily constructed basis comes. However, the computational costs of this test are prohibitive even for small lattices, so we will not perform it here. We can, however, note that the constructed truncation with $N=16$ (the last one plotted) slightly outperforms the ``SDP2'' truncation from the previous section. The basis of the ``SDP1'' truncation is (in terms of Pauli strings) $\{XIII, XXII, ZZII\}$ and all translations. The truncation ``SDP2'' adds to that $\{YYII, IYYI, IIYY, YIIY\}$, whereas the greedily constructed $N=16$ truncation adds $\{IYYI, ZIIY, YZII, IIYZ\}$. The ground state energies at $\mu = 0.7$ are:
\begin{align*}
E_{0,\mathrm{exact}} &= -4.747924\ldots\\
E_{0,\mathrm{greedy}} &= -4.747926\ldots\\
E_{0,\mathrm{SDP2}} &= -4.752392\ldots\text.
\end{align*}
Thus, while ``SDP2'' does better than one part in $10^2$, the greedily constructed basis does far better, at one part in $\sim 10^6$. Note that because the SDP solver can have substantial numerical errors, the performance of the $N=16$ truncation is consistent with it yielding the exact ground state energy for $\mu \gtrsim 0.6$.

\section{Infinite-volume limit}\label{sec:limit}

So far we have worked only with finite systems. Typically, when studying a field theory, we are interested primarily in the infinite-volume limit. Direct diagonalization and lattice Monte Carlo methods require a finite-volume extrapolation to be performed. However, the SDP-based method described here, at any fixed truncation, does not require enumerating the full Hilbert space, and so can be performed directly in the infinite-volume limit. In other words, the process of lifting the truncation already serves the role of going to the infinite-volume limit --- there is no need to take two limits independently.

To accomplish this, we must minimize the energy density instead of the energy, which is of course undefined in the infinite-volume limit. The Hamiltonian density can be written in several ways; we will use:
\begin{equation}\label{eq:Hdens_spins}
\mathcal H_{\mathrm{spins}}(r)
= -\mu \frac{\sigma_x(r)+\sigma_x(r+1)}{2} - \sum_i J_i \sigma_i(r) \sigma_i(r+1)
\text.
\end{equation}
Enforcing translational invariance\footnote{In some systems, translational invariance is in fact spontaneously broken, so that enforcing translational invariance by hand would mean studying a false vacuum. In the ferromagnetic Heisenberg spin chain this is not the case, and we won't worry about this possibility.}, the ground state energy density is just the expectation value $\langle \mathcal H_{\mathrm{spins}}(0)\rangle \equiv \epsilon_0$. The SDP is constructed along nearly the same lines as in the previous section. The quantity to minimize is the linear combination of $5$ expectation values given by $\langle \mathcal H_{\mathrm{spins}}\rangle$, and the operator basis is constructed from the set of infinite-length Pauli strings. Translational invariance is simple to implement: any two Pauli strings that appear in the problem, that are translations of each other, are forced to have equal expectation value.

The minimum number of operators needed to create a nontrivial lower bound to the ground state energy density is now just $3$: $\sigma_x(0) + \sigma_x(1)$, $\sigma_x(0) \sigma_x(1)$, and $\sigma_z(0)\sigma_z(1)$. Because our basis is constructed entirely from Pauli strings, this is split into $4$ operators, treating $\sigma_x(0)$ and $\sigma_x(1)$ separately.

That this method converges, in the limit where all operators are included, to the true ground state energy density of the infinite spin chain, follows from the discussion on convergence in the finite volume case from section~\ref{sec:spins} above. With a finite number of operators included, the infinite-volume SDP can be viewed as a finite-volume SDP for a spin chain with open boundary conditions. The desired infinite-volume energy density can be defined as the limit of those finite-volume densities.

Now that we are performing calculations directly in the infinite-volume limit, it is no longer practical to compare the SDP formulation to direct diagonalization. Tensor networks~\cite{Orus:2013kga} --- specifically matrix product states (MPSs) --- perform well in one-dimensional models, particularly the Heisenberg spin chain. Therefore, we use an MPS ansatz (detailed below) as our basis for comparison in the infinite-volume limit.

This comparison also serves to demonstrate a strategy hinted at in the introduction. The solution to the SDP formulated above, no matter what truncation is chosen, corresponds to a lower bound on the vacuum energy density. This is a nice ``dual'' to the usual variational principle, which yields an upper bound on the energy. Ordinarily, when using a variational method, the practitioner must compare the performance of several ans\"atze in order to estimate how far away the true ground state energy might be. With variational principles from both above and below, this is no longer necessary: the gap between the estimates is a reliable upper bound on the remaining error.

We choose as our ansatz a matrix product state. In the case of a spin chain, an MPS is specified by $2$ complex matrices $A_0$ and $A_1$, each of dimension $D$. The MPS $|\Psi\rangle$ is defined by its inner products with the $Z$-basis of states, each labelled by a string of $L$ bits:
\begin{equation}
\langle s_0 s_1 \cdots s_L |\Psi\rangle = \Tr A_{s_0} A_{s_1} \cdots A_{s_L}
\text.
\end{equation}

Expectation values with respect to this state can be evaluated in time linear in $L$ and cubic in the ``bond dimension'' $D$. Because the algorithm is linear in $L$, we can easily take $L$ to be large enough that finite-volume effects are negligible. Below, we adopt $L=25$.

\begin{figure}
\centering
\includegraphics[width=0.65\textwidth]{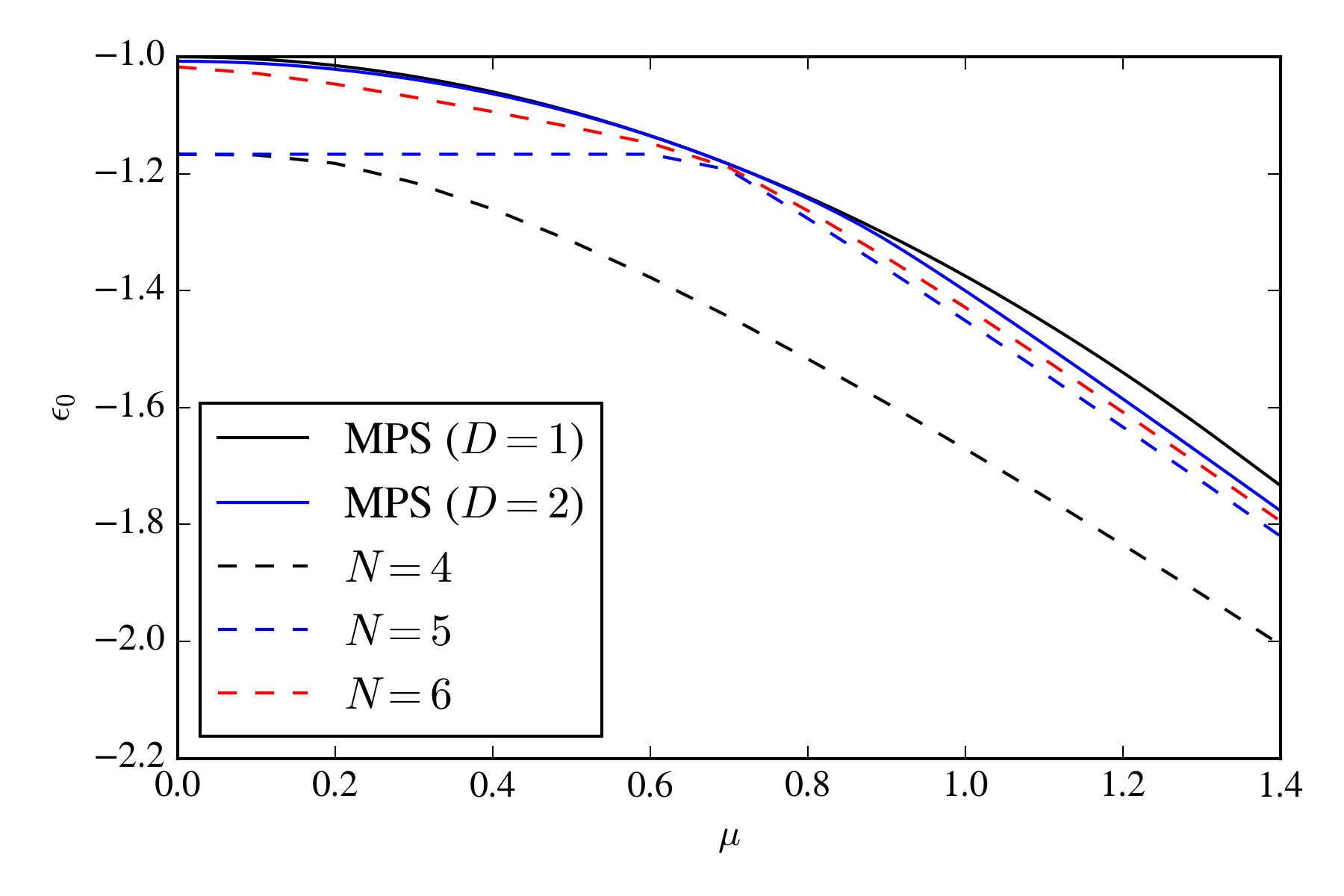}
\caption{\label{fig:limit}Energy density of the XYZ spin chain (\ref{eq:H_spins}). Upper bounds are computed with matrix product states of bond dimension $D=1$ and $D=2$; lower bounds are computed via SDP with operator truncations of size $4$, $5$, and $6$.}
\end{figure}

Figure~\ref{fig:limit} demonstrates the combination of the variational (from MPS) upper bound and SDP-obtained lower bound on the ground state energy, in the infinite-volume limit. The operators used in the $N=4$ truncation are $ IIXI $, $ IXII $, $ IXXI $, and $ IZZI$, where an infinite string of $I$s before and after are implicit in the Pauli strings. Operators added to this basis for the $N=5$ and $N=6$ truncations are determined automatically according to the procedure described in the previous section, working again at $\mu=0.7$. The first operator to be added is $IYYI$, and for the $N=6$ truncation we add $ZIZI$.

\section{Scalar field theory}\label{sec:scalar}

As a final demonstration, we turn to field theory consisting of one scalar field with an interaction term proportional to $\phi^4$. On the lattice, the Hamiltonian describing the system is
\begin{equation}\label{eq:H_scalar}
H_{\mathrm{scalar}} = \sum_{\langle r r'\rangle} \frac{\left(\phi(r) - \phi(r')\right)^2}{2} + \sum_r \left[\frac 1 2 \pi(r)^2 + \frac {m^2} 2 \phi(r)^2 + \lambda \phi(r)^4\right]
\text.
\end{equation}
For $\lambda = 0$, this is a free theory of scalar particles, describing continuum scalar field theory as $m^2 \rightarrow 0$. We will not bother with taking careful continuum limits here, using $m^2 = 0.04$ in the main examples below.

As in section~\ref{sec:limit}, it is convenient to work directly in the infinite volume limit, assuming and enforcing translational invariance to connect expectation values at different sites. We define a Hamiltonian density operator
\begin{equation}\label{eq:H_scalar_dens}
\mathcal H_{\mathrm{scalar}}(r)
=
\frac 1 2 \pi(r)^2
+
\frac 1 2 \left(\phi(r) - \phi(r+1)\right)^2
+
\frac{m^2}2 \phi(r)^2
+
\lambda \phi(r)^4
\text.
\end{equation}
It is the expectation $\langle \mathcal H_{\mathrm{scalar}}(0)\rangle$ that is to be the minimization target of the SDP.

Lattice scalar field theory can be thought of as an infinite set of coupled anharmonic oscillators. As a result, the SDP formulation has much in common with that of section~\ref{sec:anharmonic}. As with the spin chain, a complete basis of operators can be organized into the $1$-point functions, $2$-point functions, and so on, with a sensible physical heuristic being that dropping higher $n$-point functions has a small effect on the bound achieved. However, now each \emph{local} operator is of the form $\phi^a(r) \pi^b(r)$ for some non-negative integers $a$ and $b$.
As with the anharmonic oscillator, a complete basis of operators must be infinite even on a single site. A minimal truncation (any smaller yields no bound on $\langle \mathcal H_{\mathrm{scalar}}(0)\rangle$) consists of $\{\phi(0), \phi(1), \pi(0), \phi(0)^2\}$.

\begin{figure}
\centering
\includegraphics[width=0.65\linewidth]{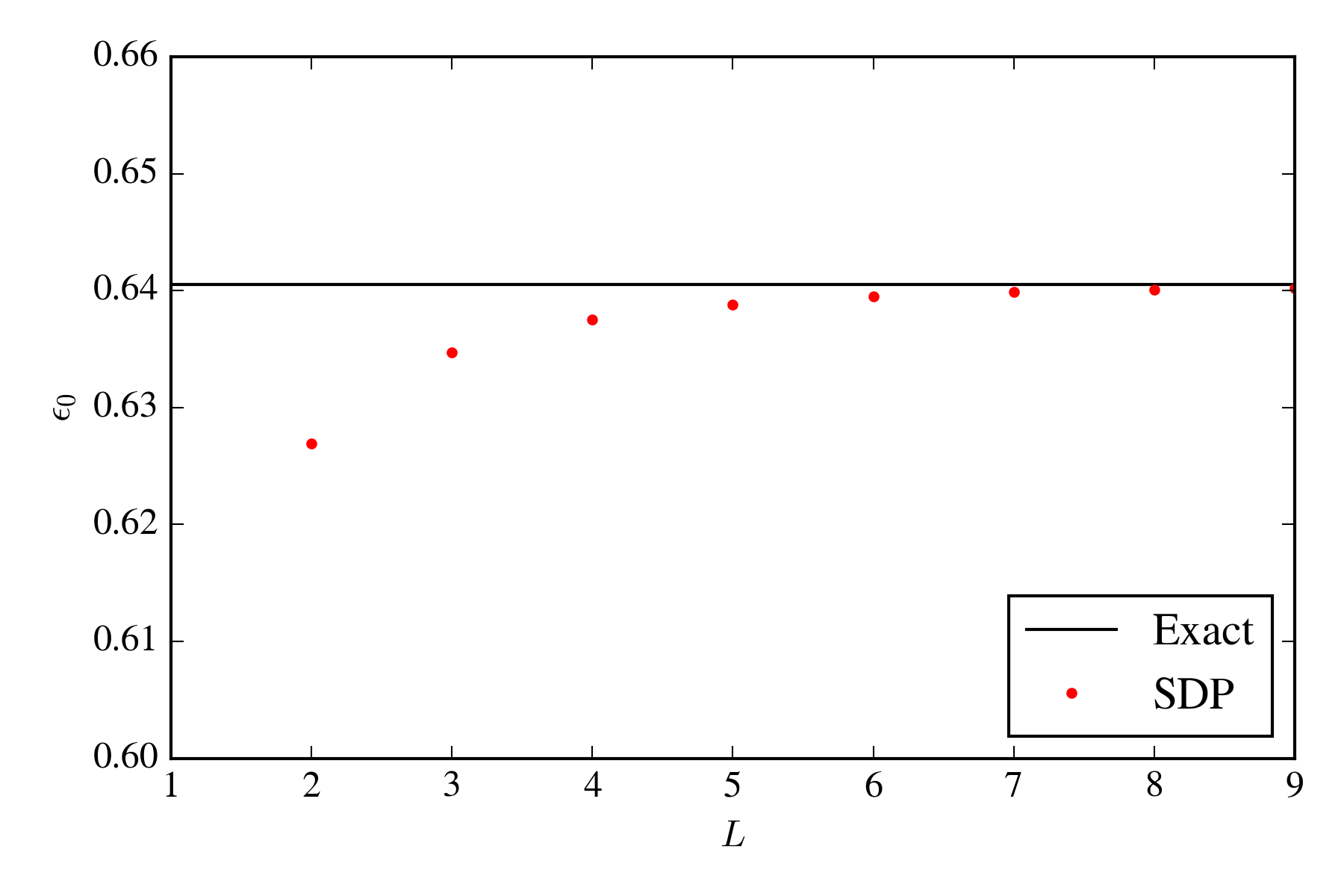}
\caption{Bounds on the vacuum energy density of lattice scalar field theory with $m = 0.1$ and $\lambda = 0$. The parameter $L$ sets the separation of the largest two-point function to play a role in the bound. As $L\rightarrow \infty$, the bound converges to the exact value.\label{fig:scalar-free}}
\end{figure}

It is instructive to begin with the free Hamiltonian; that is, (\ref{eq:H_scalar}) with $\lambda = 0$. Here the minimal basis shrinks to $\{\phi(0), \phi(1), \pi(0)\}$ --- this yields a lower bound, but not the exact vacuum energy density. The exact value can be found by rewriting the Hamiltonian in momentum basis, where it reads as an integral over uncoupled harmonic oscillators:
\begin{equation}
H_{\mathrm{free}} = \int dk\;\left[\frac{\Pi(k)^2}{2} + \left(m^2 + 4 \sin^2 \frac k 2\right) \frac{\Phi(k)^2}{2}\right]
\end{equation}
with $\Pi(K)$ and $\Phi(k)$ the canonically conjugate momentum and position operator for the mode of momentum $k$.

From this form of the Hamiltonian, it is clear that a truncation consisting only of the operators $\Phi(k)$ and $\Pi(k)$ is sufficient to yield the exact vacuum energy density. This is equivalent (after a Fourier transform) to including all one-point functions $\phi(r)$ and $\pi(r)$. Using translational invariance, this can be reduced to $\pi(0)$ alone, and $\phi(r)$ for all sites $r$. This exact truncation is still infinite. In practice, we are reduced to approximations obtained by limiting $|r| < L$ for some finite $L$. Figure~\ref{fig:scalar-free} shows how the bound compares to the exact result as $L$ is increased. Note that the run-time of the minimization is polynomial in $L$.

To address the interacting theory we begin with the minimal truncation, mentioned above, of $\{\phi(0), \phi(1), \pi(0), \phi(0)^2\}$, and add the translation of this basis by $L$ sites in either direction, as well as operators $\phi(0)^3,\ldots,\phi(0)^N$. This defines a two-parameter family of truncations, of size linear in both $L$ and $N$. There is no reason to believe that this set of truncations converges; that is, that the limit as $L,N\rightarrow \infty$ coincides with the true ground state energy density. Nevertheless it appears to perform well for a wide range of couplings.

\begin{figure}
\hfil
\includegraphics[width=0.47\textwidth]{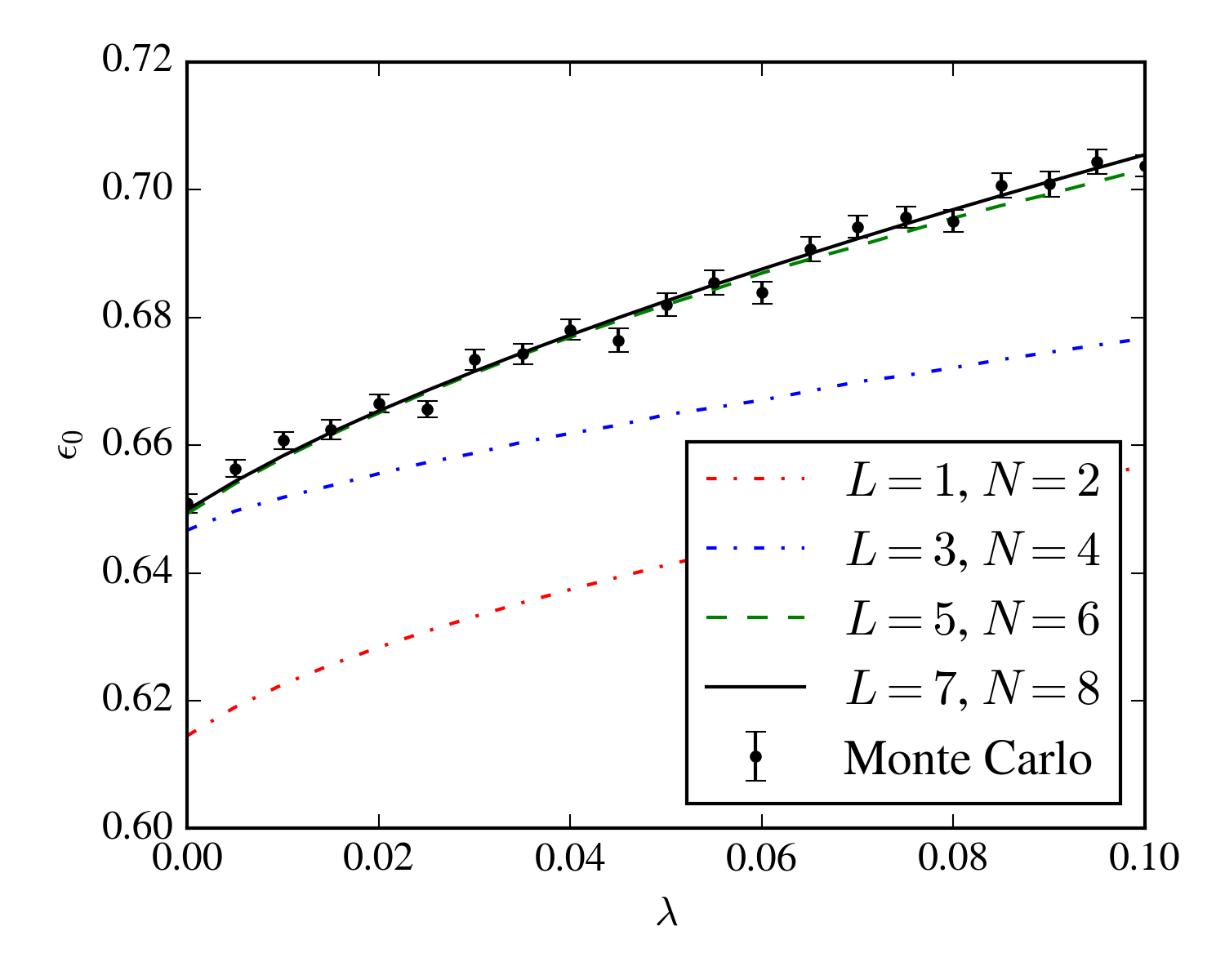}
\hfil
\includegraphics[width=0.47\textwidth]{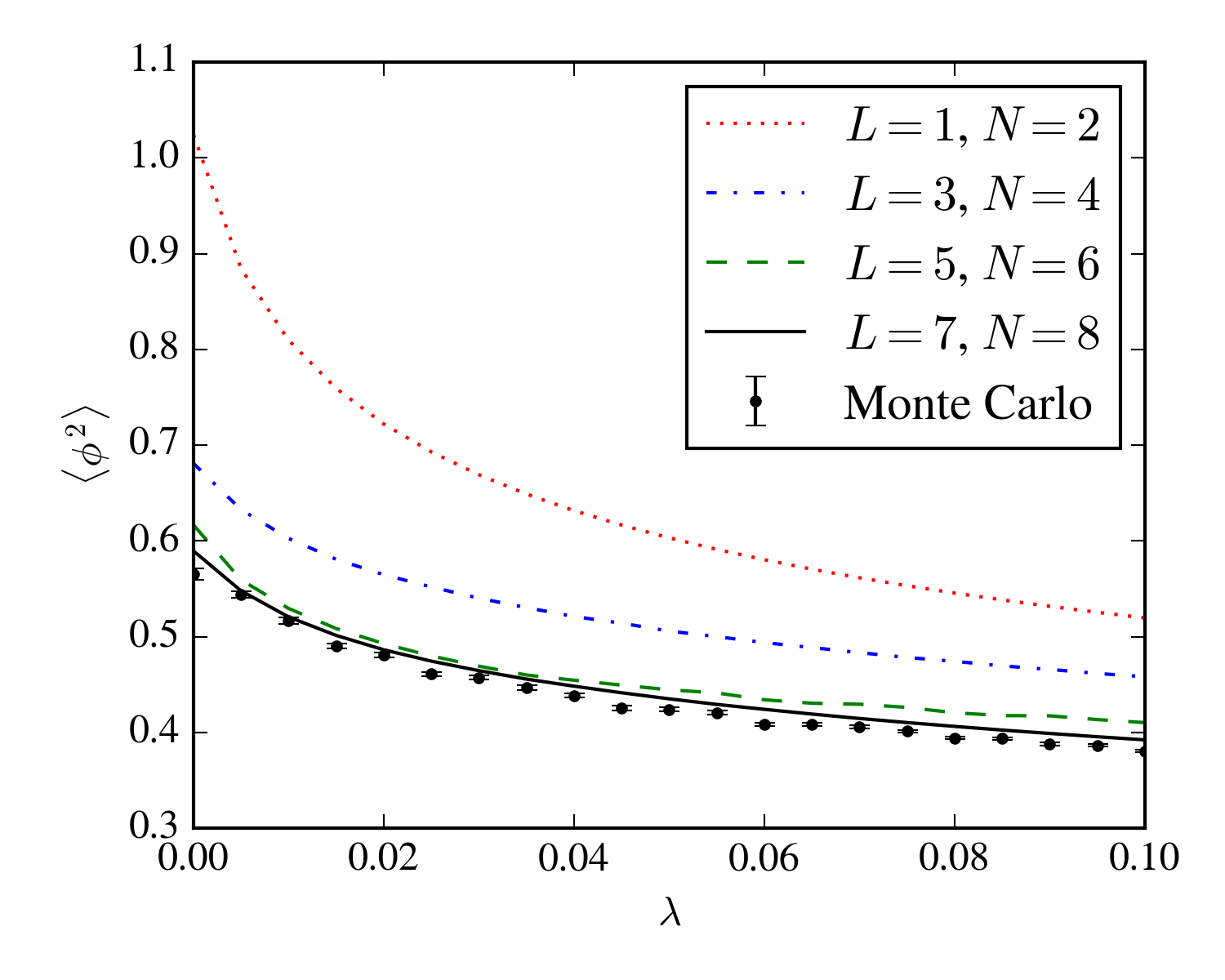}
\hfil
\caption{\label{fig:scalar}Vacuum energy density and expectation value of $\phi^2$ for lattice scalar field theory, in the infinite volume limit, as a function of the $\phi^4$ coupling $\lambda$. The energy density (left) and expectation $\langle \phi^2\rangle$ (right) are compared with a Monte Carlo calculation run on a $1500 \times 30$ lattice, large enough for finite volume and temperature effects to be negligible. All calculations are done at $m = 0.2$.}
\end{figure}

The left panel of figure~\ref{fig:scalar} shows the results of this procedure for a range of couplings $\lambda$, computing bounds on the vacuum energy density. As usual, the solution of the SDP also yields expectation values; the right panel of the same figure shows a corresponding estimate of $\langle \phi^2 \rangle$. Both are compared against unbiased estimates from a lattice Monte Carlo calculation, performed on an anisotropic lattice (to approach the Hamiltonian limit) with $1500 \times 30$ sites, with a time-spacing of $0.02$, corresponding to a volume of $30 \times 30$ in lattice units.

As with the anharmonic oscillator, numerical instabilities begin to appear once the order of the local operators reaches or exceeds $\phi^{10}$.

\section{Discussion}\label{sec:discussion}

The study of ground state properties of lattice field theories can be formulated as an SDP --- this is an unsurprising corollary of the work of~\cite{Han:2020bkb,Berenstein:2021dyf,Berenstein:2021loy} formulating quantum mechanical eigenproblems as SDPs. We have seen that this approach yields quantitatively precise results for both spin chains and lattice scalar field theory, at much lower computational cost than would be required by a direct diagonalization. Furthermore, because the SDPs yield lower bounds on the ground-state energy, and standard variational methods yield upper bounds, the two can be combined to rigorously estimate the ground state energy, with precisely known systematics. This family of methods is fairly new, and a great deal of work remains to be done to understand the capabilities and limitations of SDP-based computations.

Particularly in the context of quantum field theories, the ground state energy by itself is not a sensible physical quantity. The ground state energy can be trivially changed by adding a constant to the Hamiltonian, which does not affect any meaningful physical observable. Moreover, in the continuum limit the vacuum energy density diverges, and cannot be defined at all without choosing a renormalization prescription. In section~\ref{sec:anharmonic} we demonstrated that expectation values of arbitrary operators, not just the energy density, converge to the true value as the SDP truncation is lifted; however, these expectation estimates can no longer be interpreted as true bounds. Another possible approach, suggested in section~\ref{sec:limit}, is to use a variational method to obtain an upper bound on the ground state energy. \emph{Differences} between energy densities are physically meaningful, so that an upper bound and a lower bound (for different parameters) can be combined to give a physically relevant bound.

The solution of the SDPs corresponding to both the anharmonic oscillator and scalar lattice field theory calculations suffered from numerical instabilities that prevented higher precision, or higher-order expectation values from being studied. Precision issues in the conformal bootstrap ultimately necessitated the development of specialized SDP solvers~\cite{Simmons-Duffin:2015qma}; whether this is necessary or useful for lattice field theory studies remains to be seen. In the context of section~\ref{sec:anharmonic}, this is most likely connected to the observation of~\cite{Berenstein:2021dyf} that the positivity of the Hankel matrix $X$ requires high precision to verify.

The range of applicability of this method would be substantially improved if more than just ground-state correlation functions were available. In the context of quantum mechanical systems, it is natural to investigate excited states, as was done in~\cite{Berenstein:2021dyf,Berenstein:2021loy}. In a field theory, it is often more interesting to prepare thermal mixed states. Lattice Monte Carlo methods yield thermal information in a particularly natural way; it remains to be seen if this can be easily done with SDP-based methods.

In this work we have not investigated any issues related to renormalization or the continuum limit. Section~\ref{sec:limit} showed how calculations can be performed directly in the infinite-volume limit. This is an unusual feature for numerical methods (although more typical of analytical approaches). Optimistically, methods like those described here may be able to compute bounds directly in the continuum limit as well, by working directly with renormalized operators rather than their lattice counterparts.

The SDP formulation naturally lends itself to the addition of linear constraints on expectation values. The task of investigating finite-density states has this form: we can force the SDP solver to only look at collections of expectation values obeying $\langle n \rangle = n_0$. In the context of fermionic systems, this is an important open problem. Lattice methods are typically forbidden by the fermion sign problem from accessing regions of finite fermion density in practice. Successful application of SDP-based methods to fermionic systems would allow the \emph{ab initio} study of many new physical regimes.

Finally, although here we have looked almost exclusively at field theories, ground state energies are by themselves relevant for the study of many-bound bound states. Steps in this direction were made in~\cite{Berenstein:2021loy}; here, we should point out that removing the nonlinear constraints makes the algorithm more efficient while providing no handicap in the calculation of a binding energy.

\acknowledgments
I am deeply indebted to Frederic Koehler for much guidance regarding the construction and use of SDPs. Frederic Koehler, Henry Lamm, and Brian McPeak all provided useful comments on a earlier versions of this manuscript. This work was supported by the U.S. Department of Energy under Contract No. DE-SC0017905.

\bibliographystyle{JHEPmod}
\bibliography{References}

\end{document}